\documentclass[10pt,conference,romanappendices]{IEEEtran}
\IEEEoverridecommandlockouts
\usepackage{amsmath,amssymb,epsfig,psfrag}
\usepackage[boxed]{algorithm}
\usepackage{cite}
\usepackage{algorithmic}
\newcommand{\Err}[2]{e^{#2}(#1)}
\newcommand{\headx}{X_+}
\newcommand{\tailx}{X_0}

\newcommand{\bdot}{{\mathbf \cdot}}

\newcommand{\Nx}{N_x}
\newcommand{\Ny}{N_y}

\newcommand{\beps}{\varepsilon}

\newcommand{\R}{{\mathbb R}}

\newtheorem{theorem}{Theorem}[section]
\newtheorem{lemma}{Lemma}[section]

\newtheorem{definition}{Definition}[section]

\begin{document}

\title{A Simple Message-Passing Algorithm\\ for Compressed Sensing} 

\author{Venkat Chandar, Devavrat Shah, and Gregory
 W.  Wornell\thanks{This work was supported in part by NSF under Grant
  No.~CCF-0635191, and by a grant from Microsoft Research.}
 \\
Dept.\ EECS, MIT, Cambridge, MA 02139\\
 \{vchandar,devavrat,gww\}@mit.edu}

\maketitle

\begin{abstract}
We consider the recovery of a nonnegative vector $x$ from
measurements $y = Ax$, where $A \in \{0,1\}^{m \times n}$.
We establish that when $A$ corresponds to the adjacency matrix of a
bipartite graph with sufficient expansion, a simple message-passing
algorithm produces an estimate $\hat{x}$ of $x$ satisfying $\| x -
\hat{x}\|_1 \leq O(\frac{n}{k}) \| x - x^{(k)}\|_1$, where $x^{(k)}$
is the best $k$-sparse approximation of $x$.  The algorithm performs
$O(n(\log(\frac{n}{k}))^2\log(k))$ computation in total, and the number of
measurements required is $m=O(k\log(\frac{n}{k}))$.  In the special
case when $x$ is $k$-sparse, the algorithm recovers $x$ exactly in
time $O(n\log(\frac{n}{k})\log(k))$.  Ultimately, this work is a
further step in the direction of more formally developing the broader
role of message-passing algorithms in solving compressed sensing
problems.
\end{abstract}

\section{Introduction}

Recovery of a vector $x$ from measurements of the form $y=Ax$ has been
of central interest in the compressed sensing literature.  When
restricted to binary vectors, this has been of interest in the context
of binary linear error-correcting codes.  In essence, both desire a
matrix $A$ and an estimation or decoding algorithm that allows for
faithful recovery of $x$ from $y$.  In this paper, we study the
performance of a message-passing recovery algorithm when the matrix
$A$ corresponds to the adjacency matrix of a bipartite graph with good
expansion properties.  Results of a similar flavor are well-known in
the context of coding, but have only begun to be explored in the
context of compressed sensing.

As background, there is now a large body of work in compressed
sensing.  Both \cite{donoho} and \cite{C1,C2} proposed using linear
programming (LP) to find the sparsest solution to $y=Ax$.  Since then,
many algorithms have been proposed
\cite{C14,C13,C4,C8,C10,C12,C6,C11,indyk1,IR}---see, e.g., \cite{IR}
for a summary of various combinations of measurement matrices and
algorithms, and their associated performance characteristics.  Most
existing combinations fall into two broad classes.  In the first
class, inspired by high-dimensional geometry, the measurement matrix
$A$ is typically dense (almost all entries nonzero), and recovery
algorithms are based on linear or convex optimization.  The second
class consists of combinatorial algorithms operating on sparse
measurement matrices ($A$ typically has only $O(n)$ nonzero entries).
Examples include the algorithms of \cite{C11,IR,indyk1}. In
particular, Algorithm 1 from \cite{C11} can be viewed as essentially
the Sipser-Spielman message-passing algorithm \cite{spielman}. The
algorithm we consider in this paper also falls into the second class,
and is a minor variant of the algorithm proposed in
\cite{prabhakar}. Very recent work on the use of a message-passing
algorithm to identify compressed sensing thresholds appears in
\cite{mon1,mon2}. Relative to the present paper, \cite{mon1} and
\cite{mon2} are more general in that arbitrary (i.e., even dense)
matrices $A$ are considered.  However, \cite{mon1,mon2} restrict
attention to a probabilistic analysis, while we perform an adversarial
analysis, and thus we are able to provide deterministic
reconstruction guarantees for arbitrary (nonnegative) $x$.

On the coding theory side, Gallager introduced a class of binary
linear codes known as low-density parity check (LDPC) codes, and
proposed a computationally efficient message-passing algorithm for
their decoding \cite{gallager}.  Since then, an enormous body of work
has analyzed the performance of message-passing algorithms for
decoding such codes.  In particular, \cite{spielman} showed that when
the parity check matrix of an LDPC code corresponds to the adjacency
matrix of a bipartite graph with sufficient expansion, a bit-flipping
algorithm can correct a constant fraction of errors, even if the
errors are chosen by an adversary.  In \cite{miller}, this result is
extended by showing that a broad class of message-passing algorithms,
including common algorithms such as so-called ``Gallager A'' and ``B''
also correct a constant fraction of (adversarial) errors when there is
sufficient expansion.  Finally, \cite{feldman1} suggested decoding
LDPC codes via LP, and \cite{feldman2} proved
that this LP decoder can correct a constant fraction of (adversarial)
errors when there is sufficient expansion.  We show that similar
techniques can be used to analyze the performance of the
message-passing algorithm proposed in this paper.

The contribution of this paper is the adversarial analysis of a simple
message-passing algorithm for recovering a vector $x \in R_+^n$ from
measurements $y = Ax \in R_+^m$, where $A \in \{0,1\}^{m \times n}$.
Our first result concerns exact recovery in the case that $x$ has at
most $k$ nonzero entries.  We show formally that when $A$ corresponds
to a bipartite graph with expansion factor greater than $0.5$, the
message-passing algorithm recovers $x$ exactly.  Choosing an
appropriate expander, we find that the message-passing algorithm can
recover $x$ from $O(k\log(\frac{n}{k}))$ measurements in time
$O(n\log(\frac{n}{k})\log(k))$.  Compared to the Sipser-Spielman
algorithm \cite{spielman}, this algorithm requires less expansion
($0.5$ vs.\ $0.75$), but the Sipser-Spielman algorithm works for
arbitrary (i.e., not just nonnegative) vectors $x$.  Finally,
\cite{dimakis} shows that recovery of nonnegative $x$ is possible from
far less expansion, but their algorithm is significantly slower,
with a running time of $O(nk^2)$.

As our second result, on approximate recovery, we show that when $A$
corresponds to a bipartite graph with expansion factor greater than
$0.5$, the message-passing algorithm produces an estimate $\hat{x}$
with $\ell_1/\ell_1$ error guarantee $\| x - \hat{x}\|_1 \leq
O(\frac{n}{k}) \| x - x^{(k)}\|_1$, where $x^{(k)}$ is the best
$k$-sparse approximation of $x$.  The running time of the algorithm is
$O(n(\log(\frac{n}{k}))^2\log(k))$, and the number of measurements used is
$m=O(k\log(\frac{n}{k}))$.  In the regime where $k$ scales linearly
with $n$, our algorithm is faster than almost all existing algorithms,
e.g., \cite{C12,C6,C13}; the only exception is \cite{IR}, which is
faster, and stronger, in that the multiplier $O(\frac{n}{k})$ in the
$\ell_1/\ell_1$ guarantee is only $(1+\epsilon)$.  However, relative
to the algorithm of \cite{IR}, ours has the advantage of working with
a smaller expansion factor (albeit at the cost of requiring expansion
from larger sets), and is easily parallelizable.  In
addition, we believe that this message-passing algorithm may be
applicable in more general settings, e.g., providing guarantees for
recovering ``random'' vectors when the graph is not an expander, but
possesses other properties, such as large girth.


Beyond the specific results above, this work can be viewed
as a further step toward formally connecting the theory of
message-passing algorithms with that of compressed sensing, which we
anticipate being of growing importance to further advances in the
field.

\section{Problem Model}
\label{sec:model}

As our problem model, we seek to estimate a vector $x \in \R_+^n$ of
interest from observations of the form $y = Ax \in \R_+^m$, where $A =
[A_{ij}] \in \{0,1\}^{m\times n}$ is a known measurement matrix.
Associated with $A$ is the following bipartite graph $G=(X,Y,E)$.
First, $X = \{1,\dots, n\}$, $Y = \{1,\dots, m\}$ and $E = \{(i,j) \in
X \times Y : A_{ij} = 1\}$.  Next, associated with vertex $i \in X$ is
$x_i$, the $i$\/th component of $x$, and with vertex $j \in Y$ is
$y_j$, the $j$\/th component of $y$.  Further,
\begin{align*}
\Nx(i) &= \{j \in Y : (i,j) \in E\}, \quad \text{for all $i\in X$},\\
\Ny(j) &= \{i \in X : (i,j) \in E\}, \quad \text{for all $j \in Y$}.
\end{align*}
Note that the degrees of $i \in X$, $j \in Y$ are $|\Nx(i)|$,
$|\Ny(j)|$, respectively.  The structure in $A$ is specified via
constraints on the associated graph $G$.  Specifically, $G$ is a
$(c,d)$-regular $(k,\alpha)$-expander, defined as follows.
\begin{definition}[Expander]
A given bipartite graph $G=(X,Y,E)$ is a $(c,d)$-regular
$(k,\alpha)$-expander if every vertex $i \in X$ has degree $c$, every
vertex $j \in Y$ has degree $d$, and for all subsets $S\subset X$ such that
$|S|\leq k$ we have $|\Gamma(S)| \geq \alpha c |S|$, where
$\Gamma(S)\triangleq\cup_{i\in S}\Nx(i)$.
\end{definition}

\section{An Iterative Recovery Algorithm}
\label{sec:algorithm}

The message-passing algorithm for iteratively recovering $x$ is
conceptually very simple.  The algorithm maintains two numbers for
each edge $(i,j) \in E$, corresponding to a message in each direction.
Let $t \geq 0$ denote the iteration number and $m_{i \rightarrow
j}^{(t)}, m_{j \rightarrow i}^{(t)}$ denote the two messages along
edge $(i,j) \in E$ in the $t^{th}$ iteration.  The principle behind
the algorithm is to alternately determine lower and upper bounds on
$x$.  Specifically, $m_{i\to j}^{2t}$ and $m_{j\to i}^{2t+1}$ are
lower bounds on $x_i$ for all $t \geq 0$; $m_{i\to j}^{2t+1}$ and
$m_{j\to i}^{2t}$ are upper bounds on $x_i$ for $t\geq 0$.  Also,
these lower (respectively, upper) bounds are monotonically increasing
(respectively, decreasing).  That is,
\begin{equation*}
m_{i\to j}^0 \leq m_{i\to j}^2 \leq \dots; ~~m_{i\to j}^1 \geq m_{i\to
  j}^3 \geq \dots.
\end{equation*}
Formally, the algorithm is given by the following pseudocode.
\begin{algorithm}
\caption{Recovery Algorithm}
\label{alg:bailoutsub}
\begin{algorithmic}[1]
\STATE Initialization ($t = 0$):  for all $(i,j) \in E$, set $m_{i\to
  j}^0 = 0$.

\STATE Iterate for $t=1,2,\dots$:
\begin{itemize}
\item[a)] $t = t + 1$, update messages for all $(i,j) \in E$ via
\begin{align}
m_{j\to i}^{2t-2} & = y_j - \sum_{k \in \Ny(j)\setminus i} m_{k\to j}^{2t} \label{eq:1}\\
m_{i\to j}^{2t-1} & = \min_{l \in \Nx(i)} \left( m_{l \to i}^{2t-2} \right) \label{eq:2}\\
m_{j\to i}^{2t-1} & = y_j - \sum_{k \in \Ny(j)\setminus i} m_{k\to j}^{2t-1} \label{eq:3} \\
m_{i\to j}^{2t} & = \max \left[0, \max_{l \in \Nx(i)} \left( m_{l \to i}^{2t-1} \right)\right] \label{eq:4}
\end{align}
\item[b)] Estimate $x_i$ via $\hat{x}_i^{s} = m_{i\to j}^{s}$ for $s=2t-1, 2t$.
\end{itemize}
\STATE Stop when converged (assuming it does).
\end{algorithmic}
\end{algorithm}

\section{Main Results}
\label{sec:results}

Our results on the performance of the message-passing algorithm are
summarized in the following two theorems.  The first establishes that the
algorithm is able to recover sparse signals exactly.
\begin{theorem}
\label{lemma:bailoutsub}
Let $G$ be a $(c,d)$-regular 
$(\lfloor \frac{2}{1+2\epsilon}k+1\rfloor,\frac{1}{2}+\epsilon)$-expander,
for some $\epsilon > 0$.  Then, as long as
$k \geq \|x\|_0 = |\{ i\in X : x_i \neq 0\}|$, the estimate produced
by the algorithm
satisfies $\hat{x}^t = x$, for all $t \geq T = O(\log(k))$ .
\end{theorem}
The second theorem establishes that the algorithm can approximately
recover $x$ that are not exactly $k$-sparse.
\begin{theorem}
\label{lemma:robust}
Let $G$ be a $(c,d)$-regular
$(\lfloor\frac{k}{2\epsilon}+1\rfloor,\frac{1}{2}+\epsilon)$-expander,
for some $\epsilon > 0$.  Let $x^{(k)}$ denote the best $k$-sparse
approximation to $x$, i.e.,
\begin{equation*}
x^{(k)} = \min_{z \in R_+^n : \|z\|_0 \leq k} \|x-z\|_1.
\end{equation*}
Then, for all $t \geq T = O(\log(k)\log(cd))$,
\begin{equation*}
\| x - \hat{x}^t \|_1 \leq \left(1+\frac{d}{2\epsilon}\right) \|x -
x^{(k)}\|_1.
\end{equation*}
\end{theorem}
As a sample choice of parameters, it is well-known that there exist
expanders with $c=O(\log(\frac{n}{k}))$ and $d=O(\frac{n}{k})$.  With
such an expander, we use $O(k\log(\frac{n}{k}))$ measurements.  The
factor multiplying the error in the $\ell_1/\ell_1$ guarantee is
$O(\frac{n}{k})$, and the algorithm can be implemented sequentially in
$O(n(\log(\frac{n}{k}))^2\log(k))$ time, or in parallel
$O(\frac{n}{k}\log(k)\log(\frac{n}{k}))$ time using $O(n)$ processors.
In particular, when $k = \Theta(n)$---a regime typically of interest
in information-theoretic analysis---the algorithm provides a constant
factor $\ell_1/\ell_1$ guarantee with $O(n\log(n))$ running time.

\section{Analysis}

\subsection{Proof of Theorem~\ref{lemma:bailoutsub}}

We start by observing a certain monotonicity property of the messages.
For each $i \in X$, the messages $m_{i\rightarrow j}^{2t}$ are
monotonically nondecreasing lower bounds on $x_i$, and the messages
$m_{i \rightarrow j}^{2t+1}$ are monotonically nonincreasing upper
bounds on $x_i$.  This can be easily verified by induction.  Given
this monotonicity property, clearly the messages at even and odd times
have limits: if these messages are equal after a finite number of
iterations, then the algorithm recovers $x$.  We establish that this
is indeed the case under the assumptions of
Theorem~\ref{lemma:bailoutsub}.

To this end, define $W_{2t} = \{ i \in X : x_i > m_{i\to \cdot}^{2t}\}$, i.e.,
$W_{2t}$ is the set of vertices whose lower bounds are incorrect after $2t$ iterations.
Clearly, $|W_0| \leq k$ since $\|x\|_0 \leq k, $ and the lower bounds to $x_i$,
$m_{i\to \bdot}^{2t}$, are nonnegative for all $t$.  The monotonicity property of
the lower bounds implies that for $0 \leq s < t$, $W_{2t} \subseteq W_{2s}$.  Therefore,
it is sufficient to establish that $|W_{2t+2}| < (1-2\epsilon)|W_{2t}|$
if $0 < |W_{2t}| \leq k$; this implies that after $O(\log(k))$ iterations,
$W_{2t}$ must be empty.

Suppose $0 < |W_{2t}| \leq k$.  Since
$W_{2t + 2} \subset W_{2t}$, it suffices to show that at least a $2\beps$ fraction
of the vertices in $W_{2t}$ are not in $W_{2t+2}$ .  We prove
this by using the expansion property of $G$ (or matrix $A$).
 Let $V = \Gamma(W_{2t}) \subset Y$ be the set of all neighbors of $W_{2t}$.
Let $T \subset X$ be $\{i \in X : \Nx(i) \subset V\}$.  Since $G$ is
$(c,d)$-regular, $|V| \leq c |W_{2t}|$.  Also, by definition
$W_{2t} \subset T$.  We state three important properties of $T$:
\begin{itemize}
\item[P1.] $|T| < 2 |W_{2t}|/(1+2\epsilon)$.  Suppose not.  Then,
consider any $T' \subset T$ with $|T'| = \lfloor 2 |W_{2t}|/(1+2\epsilon)+1\rfloor$.
We reach a contradiction as follows:
\begin{align*}
c |W_{2t}| \geq ~|V| ~ \geq |\Gamma(T')|
           \geq |T'| (1+2\epsilon)\frac{c}{2}
           > |W_{2t}| c.
\end{align*}

\item[P2.] Let $U = \{i \in X : m_{i\to \bdot}^{2t+1} > x_i\}$.  Then, $U \subset T$.
This is because $m_{i\to \bdot}^{2t+1} = x_i$ if there exists $j \in \Nx(i)$
such that $j \notin V$.  To see this, note that for such
a $j$, all $k \in \Ny(j) \setminus i$ are not
in $W_{2t}$, and hence $x_k = m_{k \to j}^{2t}$ for all these $k$, so
$y_j - \sum_{k \in \Ny(j) \setminus i} = x_i$.

\item[P3.] Let $T^1 = \{i \in T : \exists j \in V \text{~s.t.~} \Ny(j) \cap T = \{i\}\}$.
Then, $|T^1| \geq 2\epsilon |T|$.  To see this, let $A = |\{ j \in V : |\Ny(j) \cap T| = 1\}|$,
and let $B = |V| - A$.  Then, number of edges between $T$ and $V$ is at least $2B + A$, and
since $G$ is $(c,d)$-regular, the number of edges between $T$ and $V$
is at most $c|T|$.  Therefore, $A+ 2B \leq c|T|$.   Now, by
[P1], $|T| < 2k/(1+2\epsilon)$, so
$|\Gamma(T)| ~\geq c |T| (1+2\epsilon)/2$.  Therefore,
$A + B \geq c |T| (1+2\epsilon)/2$, whence $A \geq 2\epsilon c|T|$.
\end{itemize}
To complete the proof, note that $T^1 \subset W_{2t}$,
and $|T^1|\geq 2\epsilon |T|$ by [P3].  For each $i \in T^1$,
let $j(i) \in V$ be its unique neighbor in the definition of $T^1$,
i.e., $\Ny(j(i)) \cap T = \{i\}$.  Then, [P2] implies that for all
$k \in \Ny(j(i)) \setminus i$, we have $m_{k\to j(i)}^{2t+1} = x_k$.
Therefore, $m_{j(i)\to i}^{2t+1} = x_i$, so
$m_{i\to \bdot}^{2t+2} = x_i$.  Thus, $i \notin W_{2t+2}$, i.e.,
$T^1 \subset W_{2t} \setminus W_{2t+2}$, completing the proof of
Theorem~\ref{lemma:bailoutsub}.


\subsection{Proof of Theorem~\ref{lemma:robust}}
\label{sec:robust}

This section establishes Theorem~\ref{lemma:robust} in
two steps.  First, using techniques similar to those used
to prove Theorem~\ref{lemma:bailoutsub}, we obtain a very weak
bound on the reconstruction error.  Next, we improve this weak bound,
by showing that when the error is large, it must be reduced significantly
in the next iteration.  This yields the desired result.

Given $x \in \R^n_+$, let $x^{(k)}$ denote the best $k$-term
approximation to $x$.  Let $\headx = \{ i \in X : x^{(k)}_i \neq 0\}$,
and let $\tailx=X \, \setminus\, \headx$.  For an arbitrary $S\subset
X$, let $\Err{S}{t}=\sum_{i\in S} (x_i\!-\!\hat{x}_i^{t})$ at the end of
iteration $t$; recall that $\hat{x}^t$ is the algorithm's estimate
after $t$ iterations.  Note that $\hat{x}^{2s}_i \leq x_i \leq
\hat{x}_i^{2s+1}$, so $\Err{S}{t}\ge0$ for even $t$, and
$\Err{S}{t}\le0$ for odd $t$.

Now, we state the first step of the proof, i.e.,
the weak bound on reconstruction error.
\begin{lemma}
\label{lemma:robust1}
Let $G$ be a $(c,d)$-regular 
$(\lfloor \frac{2k}{1+2\epsilon}+1\rfloor,\frac{1}{2}+\epsilon)$-expander,
 for some $\epsilon > 0$.  Then, after $t = O(\log k)$ iterations,
\begin{equation*}
\|x-\hat{x}^{t}\|_1 \leq O\left((cd)^{O(\log(k))}\log(k)\right) \|x-x^{(k)} \|_1.\end{equation*}
\end{lemma}
\begin{IEEEproof}
We copy the proof of Theorem~\ref{lemma:bailoutsub}.
Let $V = \Gamma(\headx)$ be the set of neighbors of $\headx$, and
let $S'=\{i \in X: \Nx(i) \subset V\}$.  Also, define sets $S_\ell, \ell \geq 0$ as follows:
\begin{equation*} S_0 = X \setminus S', ~~ S_1 = \{ i\in S': \exists j \in V \text{~s.t.~} \Ny(j)\cap S' = \{i\} \}, \end{equation*}
and for $\ell \geq 2$,
\begin{equation*} S_\ell =
\left\{ i\in S': \exists j \in V \text{~s.t.~} \Ny(j)\cap \left(S'\setminus \cup_{\ell' < \ell} S_{\ell'}\right) = \{i\} \right\}.
\end{equation*}
We note that by arguments similar to those used to establish property [P3], it
follows that $|S_{\ell}| \geq 2\beps | S'\setminus \cup_{\ell' < \ell} S_{\ell'}|$.
Also, [P1] implies that $|S'|\leq \frac{2k}{1+2\epsilon}$.
Therefore, $S_{\ell}$ is empty for $\ell\geq O(\log k)$.

Adapting arguments used in the proof of Theorem~\ref{lemma:bailoutsub}, we
bound $\Err{S_\ell}{2\ell}$ for $\ell \geq 0$.  First, by definition
$S_0 \subset \tailx$, so $\Err{S_0}{0} \leq \|x-x^{(k)} \|_1.$
Now, consider $\Err{S_1}{2}$.  By definition, each vertex $i\in S_1$
has a unique neighbor $j$, i.e., a neighbor $j$ such that
$\Ny(j) \setminus i \subset S_0$.  Therefore,
\begin{equation*}
x_i-\hat{x}_i^{2} \leq \sum_{i'\in \Ny(j)\setminus i}
(\hat{x}_{i'}^{1}-x_{i'}).
\end{equation*}
Each $i'\in S_0$, so for each $i'$ we have a neighbor $j' \not \in V$, i.e.,
$\Ny(j') \subset \tailx$.  Therefore,
\begin{equation*}
\hat{x}_{i'}^{1}-x_{i'} \leq \sum_{i'' \in \Ny(j') \setminus i'}
(x_{i''} - \hat{x}_{i''}^{0}),
\end{equation*}
where all $i''\in \tailx$.  Thus,
\begin{equation*}
x_i-\hat{x}_i^{2} \leq \sum_{i'\in \Ny(j)\setminus i} ~ \sum_{i''\in
  \Ny(j') \setminus i'} (x_{i''} - \hat{x}_{i''}^{0}),
\end{equation*}
and summing over all $i\in S_1$, we obtain
\begin{equation*}
\Err{S_1}{2} \leq \sum_{i\in S_1}\sum_{i'\in \Ny(j)\setminus
  i} ~ \sum_{i''\in \Ny(j') \setminus i'} (x_{i''}-\hat{x}_{i''}^{0}).
\end{equation*}
Now, we bound the number of times a particular vertex $i''\in S_0$
can appear on the right-hand side of the above inequality.  $i''$ can only
occur in sums corresponding to a vertex $i\in S_1$ such that there exists
a walk of length $4$ between $i''$ and $i$ in $G$.  Therefore, $i''$ can
occur in at most $(cd)^2$ terms; hence,
\begin{equation*}
\Err{S_1}{2}\leq (cd)^2\Err{S_0}{0}.
\end{equation*}

Similarly, we can bound $\Err{S_\ell}{2\ell}$ for $\ell >1$ by
induction.  Assume that for all $\ell' < \ell$,
\begin{equation*}
\Err{S_{\ell'}}{2\ell'} \leq (cd)^{2\ell'}\|x-x^{(k)} \|_1.
\end{equation*} For each vertex $i\in S_\ell$, there exists a unique
neighbor $j$, i.e., $j$ satisfies $\Ny(j)\setminus i \subset
\cup_{\ell'<\ell} S_\ell$.  Thus, $x_i-\hat{x}_i^{2\ell)}\leq
\sum_{i'\in \Ny(j)\setminus i} (\hat{x}_{i'}^{2\ell-1} - x_{i'})$.  As
before, each $i'$ has a unique neighbor $j'$, and summing over $i \in
S_\ell$, we obtain
\begin{equation*}
\Err{S_{\ell}}{2\ell} \leq \sum_{i\in S_\ell}\sum_{i'\in
  \Ny(j)\setminus i} ~\sum_{i''\in \Ny(j')\setminus i'}
(x_{i''}-\hat{x}_{i''}^{2\ell-2}),
\end{equation*}
where all $i''\in \cup_{\ell'<\ell-1}S_\ell$.  Again, each
$i''$ can only occur $(cd)^2$ times, so we conclude
that
\begin{align*}
\Err{S_\ell}{2\ell} &\leq (cd)^2\sum_{\ell'=0}^{\ell-2}
\Err{S_{\ell'}}{2\ell-2}
\leq (cd)^2\sum_{\ell'=0}^{\ell-2} \Err{S_{\ell'}}{2{\ell'}}\\
&\leq (cd)^2\sum_{\ell'=0}^{\ell-2} (cd)^{2\ell'}\|x\!-\!x^{(k)} \|_1
\leq (cd)^{2\ell'}\|x\!-\!x^{(k)}\|_1,
\end{align*}
where the second inequality is true because of the monotonicity property of
the lower bounds.  Thus, we have shown inductively that
$\Err{S_\ell}{2\ell}\leq (cd)^{2\ell}\|x-x^{(k)} \|_1$ for all $\ell$.
Since there are at most $O(\log k)$ nonempty sets $S_{\ell}$, it follows that
after $t = O(\log k)$ iterations,
\begin{align*}
\|x\!-\!\hat{x}^{t}\|_1 &\leq \sum_{\ell} \Err{S_\ell}{2\ell}
\leq O\left((cd)^{O(\log(k))}\log(k)\right) \|x\!-\!x^{(k)}\|_1.
\end{align*}
\end{IEEEproof}

On one hand, Lemma~\ref{lemma:robust1} gives a weak bound on the
reconstruction error, as the multiplier is poly($n$).  On the other
hand, it provides good starting point for us to boost it to obtain a
better bound by using the second step described next.  To that end, we
first state a definition and lemma adapted from \cite{feldman2}.

\begin{definition}
Given a $(c,d)$-regular bipartite graph $G=(X,Y,E)$, let $B(S)=\{i\in
X\setminus S: \Nx(i)\cap \Gamma(S)>\frac{c}{2}\}$ for any $S\subset
X$.  For a given constant $\delta>0$, a $\delta$-matching is a set
$M\subset E$ such that: (a) $\forall j\in Y$, at most one
edge of $M$ is incident to $j$;  (b) $\forall i \in S\cup B(S)$, at
least $\delta c$ edges of $M$ are incident to $i$.
\end{definition}
\begin{lemma}
\label{lemma:matching}
Let $G=(X,Y,E)$ be a $(c,d)$-regular $(\lfloor \frac{k}{2\epsilon}+1\rfloor,\frac{1}{2}+\epsilon)$-expander,
for some $\epsilon>0$.  Then, every $S\subset X$ of size at most $k$
has a $\frac{1}{2}+\epsilon$-matching.
\end{lemma}
To keep the paper self-contained, a proof of
Lemma~\ref{lemma:matching} is provided in Appendix~\ref{app:proof}.

We use $\delta$-matchings to prove that the reconstruction error
decays by a constant factor in each iteration.
\begin{lemma}
\label{lemma:robust2}
Let $G$ be a $(c,d)$-regular $(\lfloor\frac{k}{2\epsilon}+1\rfloor,\frac{1}{2}+\epsilon)$-expander, for some
$\epsilon > 0$.  Then,
\begin{equation*}
\Err{\headx}{2t+2} \leq \frac{1-2\epsilon}{1+2\epsilon}\Err{\headx}{2t}
+\frac{2d}{1+2\epsilon}\Err{\tailx}{2t}.
\end{equation*}
\end{lemma}

In our proof of Lemma~\ref{lemma:robust2}, we make use of the
following lemma establishing a simple invariance satisfied by the
message-passing algorithm.  Since this invariance was used earlier in
the proof of Lemma~\ref{lemma:robust1}, a proof is omitted.
\begin{lemma}
\label{lemma:witness}
For any $i \in X$, construct a set $S$ as follows.
First, choose a vertex
$j \in \Nx(i)$.  Next, for each $i' \in \Ny(j)\setminus i$, choose a vertex $w(i') \in \Nx(i')$ (note that these choices can be arbitrary).
Finally, define $S$ as $\cup_{i'\in \Ny(j)\setminus i} \Ny(w(i'))\setminus i'$.
Then, no matter how $j$ and $w(i')$ are chosen,
\begin{equation*}
x_i-\hat{x}_i^{(2t+2)} \leq \sum_{i'' \in S}
(x_{i''}-\hat{x}_{i''}^{(2t)}).
\end{equation*}
\end{lemma}

\begin{IEEEproof}[Proof of Lemma~\ref{lemma:robust2}]
Lemma~\ref{lemma:matching} guarantees the existence of a
$\frac{1}{2}+\epsilon$-matching, say $M$, for the set $\headx$ of (at
most) $k$ vertices in $X$.  We use this
$\frac{1}{2}+\epsilon$-matching to produce a set of inequalities of
the form given in Lemma~\ref{lemma:witness}.  By adding these
inequalities, we prove Lemma~\ref{lemma:robust2}.

For each $i\in \headx$, let $M(i)$ be the set of neighbors of
$i$ in the $\frac{1}{2}+\epsilon$-matching.  We construct an
inequality, or equivalently, a set $S$, for each member of $M(i)$.  We
construct the sets $S$ sequentially as follows.  Fix $i$ and $j \in
M(i)$.  For each $i' \in \Ny(j)\setminus i$, we must choose a neighbor
$w(i')$.  If $i' \in \headx$ or $i' \in B(\headx)$, set $w(i')$ to be any vertex in
$M(i')$ that has not been chosen as $w(i')$ for some previously
constructed set.
If $i' \in X\setminus
(\headx\cup B(\headx))$, choose $i'$ to be any element of
$\Nx(i')\setminus \Gamma(\headx)$ that has not been chosen as $w(i')$
for some previously constructed set.  Although it may not be
immediately apparent, we will see that this process is well-defined,
i.e., $i'$ will always be able to choose a neighbor $w(i')$ that has
not been used previously.  First, however, we complete the proof
assuming that the process is well-defined.

To that end, we establish Lemma~\ref{lemma:robust2} by adding together
all the inequalities associated with the sets $S$ constructed above.
First, consider the left-hand side of this sum.  The only terms that
appear are $x_i-\hat{x}_i^{(2t+2)}$, where $i \in \headx$, and each of
these appears at least $(\frac{1}{2}+\epsilon) c$ times since $|M(i)|
\geq (\frac{1}{2}+\epsilon) c$ for all such $i$.  On the right-hand
side, we must count how many times each term $x_i-\hat{x}_i^{(2t)}$
appears in some inequality, i.e., how many times vertex $i$ appears in
the second level of some set $S$.  We break the analysis up into two
cases.  First, assume that $i \in \headx$.  Then,
$x_i-\hat{x}_i^{(2t)}$ can appear in the second level of a set $S$
only if some vertex in $\Nx(i)$ was chosen as $w(i')$ for some $i'
\neq i$ when we were defining $S$.  This is only possible for $i' \in
\headx \cup B(\headx)$.  To bound the contribution due to such $i'$,
note that the vertices in $M(i)$ can never be chosen as $w(i')$ for
$i' \neq i$, and that every vertex in $\cup_{i \in \headx} M(i)$ is
chosen at most once.  Therefore, $x_i-\hat{x}_i^{(2t)}$ appears at
most $(\frac{1}{2}-\epsilon) c$ times.  To bound the number of
appearances of $x_i-\hat{x}_i^{(2t)}$ for $i \not \in \headx$, note
that any vertex can appear in some set $S$ at most $cd$ times.  To see
this, note that any vertex in $Y$ can appear as $w(i')$ for a set $S$
at most $d$ times, once for each of its neighbors, because a single
vertex in $X$ never chooses the same neighbor as its $w(i')$ more than
once.  The bound then follows since each vertex in $X$ has degree $c$.
Hence,
\begin{equation*}
\left(\frac{1}{2}+\epsilon\right) c\, \Err{\headx}{2t+2} \leq
\left(\frac{1}{2}-\epsilon\right)c\, \Err{\headx}{2t} +c\,d\,
\Err{\tailx}{2t},
\end{equation*}
or equivalently,
\begin{equation*}
\Err{\headx}{2t+2} \leq \frac{1-2\epsilon}{1+2\epsilon} \Err{\headx}{2t}
+\frac{2d}{1+2\epsilon} \Err{\tailx}{2t}.
\end{equation*}

Now we prove the only remaining claim that the process for
constructing the sets $S$ is well-defined.  The analysis above
implicitly establishes this already.  First, note that every $i'\in
\headx\cup B(\headx)$ has at least $(\frac{1}{2}+\epsilon)c$ distinct
neighbors that can be chosen as $w(i')$, and by definition every
$i'\in X\setminus (\headx\cup B(\headx))$ has at least $\frac{c}{2}$
distinct neighbors that can be chosen as $w(i')$.  Therefore, in order
to prove that the construction procedure for the sets $S$ is
well-defined, it suffices to show that every vertex can appear as an
$i'$, i.e., in the first level of some $S$, at most $\frac{c}{2}$
times.  For $i'\in \headx\cup B(\headx)$, at least
$(\frac{1}{2}+\epsilon)c$ of $i'$'s neighbors are in the
$\frac{1}{2}+\epsilon$-matching, so any such $i'$ appears at most
$(\frac{1}{2}-\epsilon)c$ times.  For $i'\in X\setminus(\headx\cup
B(\headx))$, by definition $\Nx(i')\cap\Gamma(\headx)\leq\frac{c}{2}$,
so any such $i'$ appears at most $\frac{c}{2}$ times.
\end{IEEEproof}

\begin{IEEEproof}[Completing proof of Lemma~\ref{lemma:robust}]
We combine lemmas \ref{lemma:robust1} and \ref{lemma:robust2}.  First,
from Lemma~\ref{lemma:robust1},  after $t=O(\log(k))$
iterations, the error satisfies the bound
\begin{equation*}
\|x-\hat{x}^t\|_1 \leq O((cd)^{O(\log(k))}\log(k)) \|x-x^{(k)} \|_1.
\end{equation*}
Lemma~\ref{lemma:robust2} implies that after an additional
$O(\log(k)\log(cd))$ iterations, the error satisfies
\begin{equation*}
\|x-\hat{x}^{t+O(\log(k)\log(cd))}\|_1 \leq
  \left(1+\frac{d}{2\epsilon}\right) \|x-x^{(k)} \|_1.
\end{equation*}
To see this, apply the inequality
\begin{equation*}
\Err{\headx}{2t+2} \leq \frac{1-2\epsilon}{1+2\epsilon} \Err{\headx}{2t}
+\frac{2d}{1+2\epsilon} \Err{\tailx}{2t}
\end{equation*}
repeatedly, and note that $\Err{\tailx}{2t}$ is monotonically
nonincreasing as a function of $t$, so $\Err{\tailx}{2t} < \Err{\tailx}{0}$.
\end{IEEEproof}

\appendices

\section{Proof of Lemma~\ref{lemma:matching}}
\label{app:proof}

The following is essentially identical to Proposition~4 and Lemma~5
in \cite{feldman2}.  We construct a $\frac{1}{2}+\epsilon$-matching by
analyzing the following max-flow problem.  Consider the subgraph of
$G$ induced by the set of left vertices $U=S\cup B(S)$ and right
vertices $V=\Gamma(S\cup B(S))$.  We assign a capacity of $1$ to every
edge in this subgraph, and direct these edges from $U$ to $V$.
Finally, we add a source $s$ with an edge of capacity
$(\frac{1}{2}+\epsilon) c$ pointing to each vertex in $U$, and a sink
$t$ with an incoming edge of capacity $1$ from every vertex in $V$.
If the maximum $s-t$ flow in this graph is $(\frac{1}{2}+\epsilon)
c|U|$, then we have constructed a $\frac{1}{2}+\epsilon$-matching.  To
see this, recall that if the capacities are integral, then the maximum
flow can always be chosen to be integral, and the edges between $U$
and $V$ with nonzero flow values in an integral maximum flow form a
$\frac{1}{2}+\epsilon$-matching.

To complete the proof, we show that the minimum $s-t$ cut in the max-flow
problem constructed above is $(\frac{1}{2}+\epsilon) c|U|$.  To see this, consider an arbitrary
$s-t$ cut $s\cup A\cup B$, where $A \subset U$ and $B\subset V$.  The capacity
of this cut is $(\frac{1}{2}+\epsilon) c (|U|-|A|)+|B|+C$, where $C$ is the number of edges
between $A$ and $V-B$.  Assume that $\Gamma(A)\not\subset B$.  Then, from
the above formula it follows that we can produce a cut of at most the same
value by replacing $B$ by $B\cup\Gamma(A)$.  Therefore, without loss
of generality we can assume that $\Gamma(A)\subset B$.
Now, an argument similar to that used to prove P1
shows that $|A| \leq \lfloor \frac{|S|}{2\epsilon}\rfloor$:
$|S|\leq k$, so if $|A| \geq \lfloor \frac{|S|}{2\epsilon}+1\rfloor$,
then there exists a set of size $k'=\lfloor \frac{|S|}{2\epsilon}+1\rfloor$
with at most $c|S|+\frac{c}{2}(k'-|S|)<(\frac{1}{2}+\epsilon)ck'$
neighbors, contradicting the $(\lfloor\frac{k}{2\epsilon}+1\rfloor,\frac{1}{2}+\epsilon)$-expansion of $G$.
Therefore, $|\Gamma(A)|\geq (\frac{1}{2}+\epsilon) c|A|$,
so the min-cut has capacity at least
$(\frac{1}{2}+\epsilon) c(|U|-|A|)+(\frac{1}{2}+\epsilon)
c|A|=(\frac{1}{2}+\epsilon) c |U|$. \IEEEQEDclosed

\end{document}